\documentclass[prd,a4paper,twocolumn,showpacs,amsmath,amssymb]{revtex4}

\usepackage{graphicx,amsmath}
\usepackage{epsfig}

\def\beq{\begin{equation}}
\def\eeq{\end{equation}}
\def\bea{\begin{eqnarray}}
\def\eea{\end{eqnarray}}

\begin{document}


\title{	Probing neutron's electric neutrality with Ramsey Spectroscopy\\
	of gravitational quantum states of ultra-cold neutrons}

\author{Katharina Durstberger-Rennhofer }
\author{Tobias Jenke}
\author{Hartmut Abele}

\affiliation{	Atominstitut, Vienna University of Technology\\
		Stadionallee 2, 1020 Vienna, Austria}

\date{\today}


\begin{abstract}

We propose to test the electric neutrality of neutrons by a new technique using the
spectroscopy of quantum states of ultra-cold
neutrons in the gravity potential above a vertical mirror. The new technique is an application of Ramsey's method of separated oscillating fields to
neutron's quantum states in the gravity potential of the earth.
In the presence of an electric field $E_z$ parallel or antiparallel to the direction of the acceleration of the earth $g$, the energy of the quantum states changes due to an additional
electrostatic potential if a neutron carries a non-vanishing charge.

In the long run our new method has the potential to improve the current limit of $10^{-21} q_e$ for
the electric charge of the neutron by 2 orders of magnitude.
\end{abstract}



\maketitle

\section{Introduction}

The smallness of the neutron charge $q$ to be less than 1.8 $\times$ 10$^{-21}$ electron charges $q_e$
(90 \% C.L.) raises serious questions about charge-quantization. The
Standard Model with three generations does not have electric charge
quantization~\cite{Foot93,Foo93}, so $q$ could be anything.  In fact, charge quantization requires an additional free parameter in the standard model, see, e.g., Ref. \cite{Ign93,Ign95}, which must be determined experimentally along with the other standard model parameters (like the coupling strengths of electro-weak and strong interaction, the Higgs boson mass, etc). If this free parameter is nonzero, it induces small modifications of the electric charges. As a consequence, so called neutral particles, like neutrons, neutrinos and atoms, carry a small 'rest charge' \cite{Arv08}.
Assuming charge conservation and the validity of the CPT theorem, this parameter has to be below $3 \times 10^{-21}$ (see, e.g., \cite{Tak91}). This most stringent limit arises from the upper limit of neutron charge $q$.

There are many extensions of the
Standard Model, which lead to electric-charge
quantization~\cite{Choudhury91}. Other suggestions include higher
dimensions~\cite{Klein26}, superstrings~\cite{schw86,Dist05}, magnetic monopoles~\cite{Dirac31} and
Grand Unified Theories (GUTs)~\cite{Pati74,Georgi74,Okun84}. Since
the Standard Model value for $q$ requires extreme fine tuning, the
smallness of this value may be considered as a hint for GUTs, where
$q$ is equal to zero. But a non zero value of $q$ would eliminate the possibility of neutron-antineutron oscillations~\cite{Baldo-Ceolin94}, which is a GUT-candidate for a violation of the baryon number by $\Delta B=2$~\cite{Gla79}. \\

That the neutron is a particle having zero electric charge has been checked
by beam-deflection experiments \cite{Bau88,Bor88}, where slow neutrons with mass $m$ pass through a strong electric field perpendicular to the beam direction. If a hypothetical neutron charge $q$ was present, one would expect a deflection $y$,
\begin{equation}
	y=\frac{q^2 E_z L^2}{2 m v^2}\;,
\end{equation}
with $E_z$ being the electric field applied over the length $L$ and $v$ the neutron velocity.

The deflection apparatus of Baumann et al. \cite{Bau88} uses a multislit system with 31 slits, 30$\mu$m wide, separated by 30$\mu$m-wide absorbing zones. With a detector slit positioned on the steep slope of the intensity profile, which is assumed to be Gaussian with $2\Delta$ full width at half maximum, a beam deflection $y$ becomes noticeable by measuring the difference in counting rate for opposite directions of the applied electric field $E_z$. Assuming a deflection much smaller than the width of the profile, the uncertainty in $y$ is given by
\begin{equation}
	\sigma_y=\frac{\Delta}{\sqrt{N}}\;,
\end{equation}
where $N$ are the total neutron counts \cite{Gae82}. In order to minimize $\sigma_y$, a high count rate and a small beam profile is desired. The sensitivity of the apparatus was such that a deflection $y = (2.3 \pm 14.7)\times 10^{-10}$m was measurable for a flight path $L$ of about $9$m, an electric field of $E_z = \pm 6 \times10^6$V/m and a neutron wavelength of $\lambda=1.2 \pm 3$nm, respectively. The sensitivity is impressive and expressed in angular resolution or momentum change it gives
\begin{equation}
	\Theta=\frac{y}{L}=2 \times 10^{-10}\;.
\end{equation}
Baumann et al. derived for the charge of the neutron
\begin{equation}
	q =(-0.4 \pm 1.1)\times 10^{-21} \; q_e\;,
\end{equation}
where $q_e$ denotes the electron charge. This measurement is in agreement with the neutrality of neutrons.

Another experiment, with ultra-cold neutrons, was conducted nearly at the same time by Borisov et al. \cite{Bor88}. The lower intensity of the UCN beam was counterbalanced by the longer time the slow UCN remained in the electric field region. The intrinsic discovery potential of this experiment was
$q = 3.6\times10^{-20} \; q_e$ per day at the former UCN source of the Leningrad VVR-M reactor. During only three days of running this experiment produced the
result
\begin{equation}
	q = -(4.3 \pm 7.1)\times 10^{-20} \; q_e\;.
\end{equation}

Up to now, all experiments probing the electric neutrality of neutrons were designed as deflection experiments (see also Ref.\cite{Plonka10}).

We propose to probe neutron's neutrality by a new technique using the spectroscopy of quantum
states in the gravity potential above a vertical mirror. The new technique is an application of
Ramsey's method of separated oscillating fields \cite{Ram56} to quantum states in the gravity
potential of the earth \cite{Abe09} equipped with an electric field in the intermediate flight path region.

Energy eigenstates in the gravity potential of the earth can be probed by a new resonance spectroscopy technique, using neutrons bouncing on a horizontal mirror~\cite{Jenke2011}. In the presence of an electric field $E_z$, the energy of quantum states in the gravity potential changes due to an additional electrostatic potential if a neutron carries a nonvanishing charge $q$. Important for this method is the fact that the energy shift differs from state to state due to the properties of a Schr\"odinger wave packet in a linear potential. We measure the energy difference between two quantum states by applying an electric field $E_z$ parallel or anti-parallel to $g$. It will allow high precision spectroscopy, because ultimately the highest precision in experiments can be obtained by
measuring frequencies.

\section{Ramsey Spectroscopy of Gravitational Quantum States of Neutrons}

\subsection{Quantum states of neutrons in the gravitational and external electrostatic potential}

Let us consider the motion of ultracold neutrons with a hypothetical electric charge $q$
in a gravitational and electric field above a horizontal mirror.
We assume their forces to act in $z$-direction,
while the mirror is aligned with the $xy$-plane at $z = 0$.
The motion in $x$- and $y$-direction is free and completely decouples from that in $z$-direction.
Without the external electric field, the problem is equivalent to the quantum bouncing ball \cite{Gib75,Gea99}.
\begin{figure}[htbp]
\begin{center}
  	\includegraphics[width=0.41\textwidth]{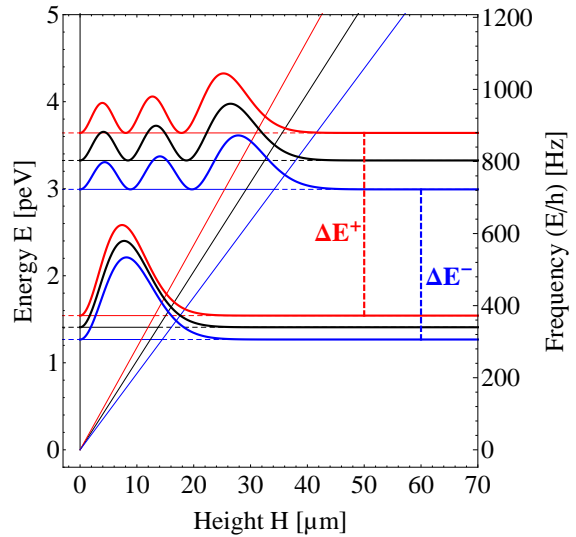}
         \caption{Energy eigenvalues and probability densities of the first and third eigenstate of a neutron in the gravitational potential of the earth (black curves). The red (blue) curves show modifications due to an electric field in parallel (antiparallel) configuration for a hypotetical neutron charge $q=5\times 10^{-16} \; q_e$.}
         \label{fig:states}
\end{center}
\end{figure}

It suffices to consider the time-dependent Schr\"odinger equation restricted to the $z$-direction
\beq
		\left\{ -\frac{\hbar^2}{2m}\frac{\partial^2}{\partial z^2} + mgz + q |\vec{E_z}| z \right\}\Psi =
		i\hbar\frac{\partial\Psi}{\partial t} \, .
	\label{eq:sdg}
\eeq
Here, $g$ is the acceleration of gravity, $m$ is the mass of the neutron and
$|\vec{E_z}|$ is the external electric field pointing in $z$-direction.
We are interested in two special cases $|\vec{E_z}|=\pm E_z$ where the electric field is oriented parallel ($+$) or antiparallel ($-$) to the acceleration of gravity.
The potential of the mirror at $z = 0$ associated with the substance of the
mirror is repulsive and much larger than the eigenenergies of the lowest
quantum states in the gravitational field. Therefore
Eq. (\ref{eq:sdg}) must be solved with the boundary condition
$\Psi(z=0,t)=0.$

The corresponding stationary Schr\"odinger equation is given by
\begin{equation}
	\left\{-\frac{\hbar^2}{2m}\frac{\partial^2}{\partial z^2}+(mg + q |\vec{E_z}|)z\right\}\;\psi_n=E_n \psi_n\;.
	\label{eq:sdgstationary}
\end{equation}
It is convenient to use rescaled units $\zeta = z/z_0$ and $\epsilon_n = E_n/E_0$ with the characteristic gravitational quantum length $z_0$ and energy scale $E_0$ of the bouncing neutron, which depend on a hypothetical electric charge of the neutron $q$:
\bea
z_0 (q) & = & \left(\frac{\hbar^2}{2m} \frac{1}{(mg + q |\vec{E_z}|)}\right)^{1/3}\\
E_0 (q) & = & \left( mg+q|\vec{E_z}| \right) z_0 (q)
	\label{z0e0}
\eea

\begin{figure*}[htb]
	\begin{center}
	\includegraphics[width=0.90\textwidth]{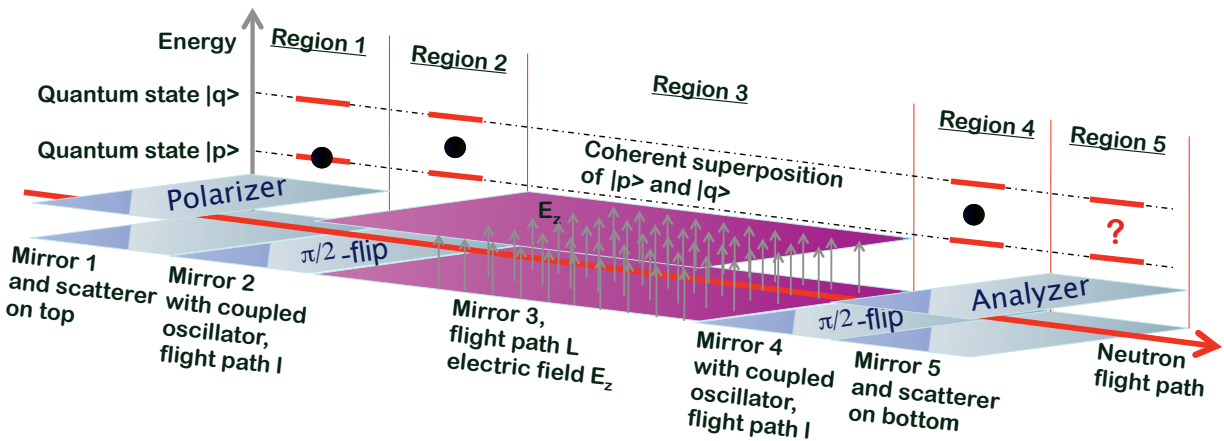}
	\caption{Proposed experimental setup. Region 1: Preparation in a specific quantum state, e.g. state one with polarizer. Region 2: Application of first $\pi/2$-flip. Region 3: Flight path with length L. Region 4: Application of second $\pi/2$-flip. Region 5: State analyzer}
	\label{fig:setup}
	\end{center}
\end{figure*}
The solutions of Eq. \eqref{eq:sdgstationary} is given in terms of Airy-functions
\begin{equation}
	\psi_n(\zeta)=N_n \, Ai(\zeta-\epsilon_n)\;,
\end{equation}
where $N_n$ is a proper normalisation factor and $\epsilon_n$ the n-th energy eigenvalue (in rescaled units).
The displacement $\epsilon_n$ of the Airy functions has to coincide with the n-th zero of the Airy function,
$Ai(-\epsilon_n)=0$, due to the boundary condition $\psi_n(0)=0$.

For zero electric charge of the neutron, the eigenenergies of the quantum bouncer are
\begin{equation}
	E_n^{(0)} = \epsilon_n mg \,z_0(q=0)\;,
\end{equation}
which gives for the lowest energy levels $E_1^{(0)} =1.41$peV, $E_2^{(0)} =2.46$peV, $E_3^{(0)} =3.32$peV. For nonzero electric charge of the neutron the energies for the two different field configurations are denoted by $E_n^\pm$.

Fig. \ref{fig:states} shows the probabiltiy densitiy of the first and third energy eigenstate (black lines) and the influence of a hypothetical electric charge $q$ of the neutron. The red (blue) curves show the eigenfunctions in presence of an electric field $+E_z$($-E_z$) in the parallel  (antiparallel) configuration, calculated for a hypothetical neutron charge of $q=5\times 10^{-16} \; q_e$.

\subsection{Ramsey's Method of Separated Oscillating Fields}

Ramsey's method \cite{Ram56}, as described for neutrons in the gravitational potential of the earth in Ref. \cite{Abe09}, probes the difference in energy shifts $\Delta E=\Delta E_q-\Delta E_p$, with $\Delta E_n=E_n^+-E_n^-$, between two levels $p$ and $q$.
We modify this experimental setup of Ramsey's resonance method for neutron's gravity states such that it is suitable for a measurement of a hypothetical charge of the neutron.
A sketch of a modified setup to test neutron's neutrality is shown in Fig. \ref{fig:setup}.

To implement Ramsey's method, one has to realize
\begin{enumerate}
	\item a state selector or polarizer
	\item a region, where one applies a $\pi/2$ pulse creating the superposition of the two states, whose energy difference should be measured,
	\item a region, where the phase evolves,
	\item a second region to read the relative phase by applying a second $\pi/2$ pulse, and finally
	\item a state detector or analyzer.
\end{enumerate}

In region 1, neutrons are prepared in a specific quantum state $\lvert p\rangle$ in the gravity potential following the procedure demonstrated in \cite{Nes02}. Above a polished mirror a rough absorbing scatterer is mounted which selects only the ground state and absorbs or scatters out higher unwanted states, see \cite{Wes07a}.

In region 2 of length $l$, the first of two identical oscillators is installed. Here, transitions between
quantum states $\lvert p\rangle$ and $\lvert q\rangle$ are induced. The oscillator frequency at resonance for a transition between states with energies $E_q$ and $E_p$ is $\nu_{pq}=(E_q-E_p)/h$ which gives for the transition $\lvert 1\rangle \rightarrow \lvert 3\rangle$ a frequency of $\nu_{13}=461.9$Hz. Driven at resonance ($\nu = \nu_{pq}$), the oscillator brings the system into a coherent superposition of the state $\lvert p\rangle$ and $\lvert q\rangle$; a $\pi/2$-pulse creates an equal superposition. The oscillator system is realized either by using oscillating magnetic gradient fields or by vibrating mirrors where a modulation of the mirror potential in height takes place.

\begin{figure}[htbp]
\begin{center}
  	\includegraphics[width=0.41\textwidth]{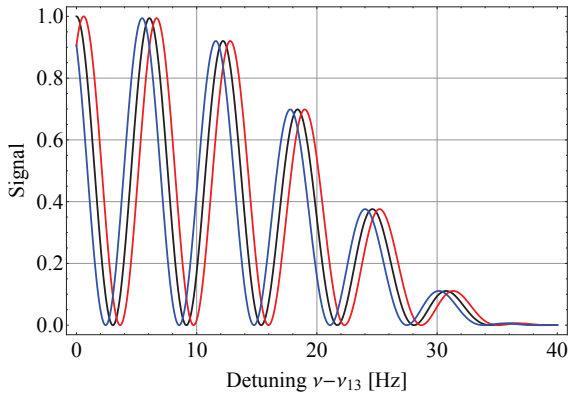}
         \caption{Expected Ramsey fringe pattern for ultracold neutrons traversing the system (black). If the neutron carries a charge of $q = 5\cdot10^{-18} \; q_e$, the detected signal will change in dependance of the direction of the applied electric field (parallel to gravity: red curve, anti-parallel: blue curve). For the calculation of this plot, the geometric parameters of Ref. \cite{Abe09} were used.}
         \label{fig:fringe}
\end{center}
\end{figure}
In the intermediate region 3, a non-oscillating mirror with a neutron flight path of $L$ and flight
time $T$ follows. It might be convenient to place a second mirror on top of the bottom mirror at a
certain height $H$. The mirrors are rounded off and are coated with gold for electrical conductivity. Field strengths of about $6\times 10^6$V/m are used.

In region 4, a second oscillator identical to region 2 in phase with the oscillator in region 2 is placed.

In section 5, the accumulated phase shift can be measured by transmission through a second state selector.

By tuning the oscillation frequency $\nu$, a typical Ramsey fringe pattern as shown in Fig. \ref{fig:fringe} (black line) will be observed. To test neutron's neutrality, three different configurations are used: the electric field is put on, thereby pointing either parallel, $+E_z$, or antiparallel, $-E_z$, to gravity, or the electric field is put off.
In the parallel and antiparallel configuration Ramsey's method probes an energy difference of $\Delta E^\pm=E_q^\pm-E_p^\pm$ (see also Fig. \ref{fig:states}).

If the neutron carries an electric charge, the frequency shift between the two resulting Ramsey fringe patterns for the parallel and anitparallel configuration will correspond to
\begin{equation}
	\Delta \nu = \nu_{pq}^{(0)}\cdot \left( \sqrt[3]{\left(1+\frac{q E_z}{mg}\right)^2}- \sqrt[3]{\left(1-\frac{q E_z}{mg}\right)^2}\right).
	\label{eq:shift}
\end{equation}
This is illustrated in Fig. \ref{fig:fringe} by the blue and the red curve for a hypothetical neutron charge of $q = 5\times0^{-18} \; q_e$. The black curve corresponds to the case of the electric field switched off or the neutron charge being $q = 0$. For this figure, the geometric parameters of the setup suggested in \cite{Abe09} were used to calculate the Ramsey fringe patterns for the transition $\lvert 1\rangle \rightarrow \lvert 3\rangle$.

\subsection{Expected sensitivity}

The search for a hypothetical charge of the neutron consists of the following measurements:
Firstly, the Ramsey pattern for the transition $\lvert p\rangle$ to $\lvert q\rangle$ is recorded with sufficient statistics
to resolve the steep Ramsey fringes. Then, the frequency of the oscillators is locked to the frequency $\nu_0$ where
the Ramsey fringes give the steepest slope.
The number of neutrons for a fixed observation time $t$
for the two different possible directions of the electric field, parallel or anti-parallel to gravity,
is measured. The corresponding number of neutrons are denoted by $N^+$ and $N^-$.
For the difference of neutron counts, we expect
\begin{equation}
	\frac{N^+}{t} - \frac{N^-}{t} = \left.\frac{\partial r\left(\nu\right)}{\partial \nu}\right |_{\nu_0} \Delta \nu\;.
\end{equation}
Here, $r\left(\nu\right)$ corresponds to the Ramsey fringe pattern expressed as a countrate and $\Delta \nu$
to the frequency shift induced by the hypothetical charge of the neutron.

With the help of Eq. (\ref{eq:shift}) and a Taylor-expansion in $q$, this formula can be re-expressed:
\begin{equation}
	q = \frac{N^+ - N^-}{t} \cdot \frac{1}{\left.\frac{\partial r\left(\nu\right)}{\partial \nu}\right |_{\nu_0} \nu_0}
			\cdot \frac{3}{4}\frac{mg}{E_z}
\end{equation}
The statistical error on $q$ is given by
\begin{equation}
	\delta q = \frac{\sqrt{2} \overline{r}}{\sqrt{N}} \cdot \frac{1}{\left.\frac{\partial r\left(\nu\right)}{\partial \nu}\right |_{\nu_0} \nu_0}
			\cdot \frac{3}{4}\frac{mg}{E_z}.
	\label{eq:sens}
\end{equation}
Here, $N$ is the total number of counted neutrons, $t$ equals to the total measuring time, $\overline{r}$ is the mean count rate $\overline{r} = N/t$, and the assumption $N^+ \approx N^- \approx N/2$ was used.

To estimate the sensitivity of the suggested method, it is useful to calculate the so-called discovery potential, i.e., the statistical limit on the hypothetical charge $q$ reached in a measuring time of one single day.
To determine this discovery potential, all ingredients of eq. \ref{eq:sens} need to be estimated:

The mean rate $\overline{r}$ profits from one of the main advantages of Ramsey's Method:
as the system is self-focussing, the steep slope of the inner Ramsey fringe stays unchanged even if the transmitted neutrons
have a certain velocity distribution. From our previous experiments at the beam position UCN/PF2 at ILL,
the mean count rate can be estimated to be $\overline{r} \approx 0.1 s^{-1}$ using all neutrons with velocities $v_x$
between $3.2$m/s and $9$m/s.
The total statistics per day is given by $N = \overline{r}\cdot T = 0.1 s^{-1}\cdot 86400 s = 8640$ neutrons.

The steepest slope of the Ramsey fringe pattern is given by value $\partial r\left(\nu\right) \approx 1$ Hz frequency shift per 	$s^{-1}$ transmission change. For this calculation, the standard neutron mirror setup as proposed in \cite{Abe09} for an 	
in-flight experiment was used. Therefore, the interaction time of the neutron with the electric field is $\tau = 0.130$s.

Baumann et al. \cite{Bau88} used an electric field of $E_z = 6 \times 10^6$V/m. The distance of the electrodes was
$3$mm. The achievable electric field scales with the square-root of the distance, thus an improvement by a factor of five by
using electrodes with a distance of $100 \mu$m is possible. Measured breakdown voltages at electrode distances of $4$mm are
around $20$MV/m \cite{Kur03}, and $70$MV/m at a distance of $100\mu$m have been reported \cite{Chi79} but all figures depend
strongly on the geometry.
There are deviations, which are proportional to the electric field $E_z$. The effects of the magnetic moment $\mu$ in magnetic
stray field can be reduced by the use of mu-metal shielding (four layers). Then the effect is smaller than $10^{-23}$eV
\cite{Bak06}. The effect of Schwinger terms, $\vec E_z\times \vec v$, has been studied recently in EDM experiments. It can be
neglected at this level of accuracy, furthermore because we are using unpolarized neutrons, where the effect cancels on average.
Effects due to the electric polarizability of the neutron are also very small \cite{Abe03}.

With these parameters, the discovery potential reads
\begin{equation}
	\delta q = 8.4\times 10^{-20}\; q_e/ {\rm day}
\end{equation}
for the transition $\lvert 1\rangle \rightarrow \lvert 3\rangle$. This sensitivity may be improved by choosing higher transitions
such as $\lvert 1\rangle \rightarrow \lvert 5\rangle$ resulting in a discovery potential of $\delta q = 4.8\times 10^{-20}\; q_e/ {\rm day}$.

The neutron-charge experiment \cite{Bau88} with the best limit on the charge was performed at the cold neutron guide H18 of ILL and the full neutron spectrum of this beam was used. For this kind of experiment it has been shown
that the reachable limit for $q$ is independent of the wave length $\lambda$ as long as the neutron spectrum is proportional to $1/\lambda^5$, which is the case for the research reactor at the ILL.
To improve the limit significantly by several orders of magnitude, we can use our method with ultra-cold neutrons, because they can be stored and thus the observation time $\tau$ can be increased by 3 orders of magnitude, which would improve the limit linearly. Furthermore, new ultracold neutron sources are under development right now and the source strength density is expected to be increased by two orders of magnitude.
This results in an ultimate - statistical - discovery potential of
\begin{equation}
	\delta q = 8.4\times 10^{-24}\; q_e/ {\rm day}
\end{equation}
as a long-term goal for this method.

\begin{acknowledgments}
H.A. would like to thank R. G\"ahler for useful discussions. This work has been supported by the German Research Foundation (DFG) under Contract No. Ab128/2-1, and by the Austrian Science Fund (FWF), project no. I 531-N20 and T 389-N16 (Hertha-Firnberg position of K. D.-R.).
\end{acknowledgments}

\bibliographystyle{apsrev}

\end{document}